\documentclass[12pt]{article}
\usepackage{epsfig}

\begin{document}
{\bfseries The \vspace{0.3in} Directed Polymer --- Directed Percolation Transition}

\begin{center}
{\bfseries Ehud Perlsman and \vspace{0.2in} Shlomo Havlin}
\end{center}
The Resnick Building, Minerva Center and Department of Physics, Bar-Ilan 
University, 52900 Ramat-Gan,\vspace{0.6in} Israel.

\begin{center}
{\bfseries  ABSTRACT}
\end{center}

\vspace{0.4in} We study the relation between the directed polymer and the directed
percolation models, for the case of a disordered energy landscape where the energies
 are taken from bimodal distribution.
 We find that at the critical concentration of the directed percolation, the directed 
 polymer undergoes a transition from the directed polymer universality class to the directed
 percolation universality class. We also find that directed percolation clusters 
 affect the characterisrics
 of  the directed polymer below the critical \vspace{0.9in} concentration.

\newpage

 Several models have been developed for directed path in disordered media.
 Two well known models 
 are  directed polymer and 
 directed percolation [1,2]. While the directed polymer is based on global optimization, the
 directed percolation can be described as a local process.  The directed polymer
  is considered as a model for physical processes such as tearing or cracks [3], while 
   the directed percolation has been used to model for example invasion of low viscosity
    fluid in high viscosity[4], as well as the interface of liquids in materials 
    such as paper[5]. The relation between the two models is not yet clear and even 
    controversial, as discussed \vspace{0.3in} bellow.

The  two models of directed polymer and directed percolation can be described in
 a similar way as follows: In a square lattice which was cut along its
 diagonal and oriented as a triangle whose apex is up and the diagonal is its
 base, we assign to each bond  a random
 number. Of all the paths leading from the apex (the origin) to the base we 
refer only to those whose direction is \emph{always} down to the base. For each 
one of these paths we calculate the sum of the random numbers along it. 

In the directed polymer model, the sum of the random numbers along the path
  defines its value. We focus our interest on the path of minimal
 value (which we call ''the optimal path''), and define
 the roughness exponent
 $\nu$ by $D \sim t^{\nu}$, where D is the mean distance of the endpoint of the
 optimal path from the center of the base, t is the size of the triangle,
 and $\nu$ is the roughness exponent. In this model the random numbers are
 usually taken from a continuous distribution, and thus  the optimal path is uniquely
 defined. Huse and Henley [1] who introduced this model found
 that $\nu \cong 2/3$ and the value 2/3 is considered to be exact. Another exponent 
 which characterizes this model is $\omega$, defined by $\sigma_{E} \sim t^{\omega}
 $ where  E is the  the value 
 of the optimal path, and $\sigma_{E}$ is its standard deviation. These two 
exponents are related by the scaling
 relation $\omega = 2\nu - 1$ [1].

In the directed percolation model the random numbers are taken from a bimodal
 (0,1) distribution, and we can define a percolation cluster as the collection of 
 lattice sites which are connected by zero sum paths to the origin.
 In this model we can also define a rougheness
 exponent by $W \sim L^{\nu}$ where W is the mean width and L is the mean 
 length of a percolation cluster.  W and
 L depend on the probability to get 0 (denoted by p) by the relations 
 $W \sim (p_{c}-p)^{-\nu_{\perp}}$, $L \sim (p_{c}-p)^{-\nu_{\parallel}}$,
 where  $p_{c}$ is the critical probability $\simeq 0.6447$, $\nu_{\perp}$
 is the transverse exponent $\simeq 1.097$, and $\nu_{\parallel}$ is the longitudinal
 exponent $\simeq 1.733$ [4]. Since $\nu = \nu_{\perp}/\nu_{\parallel} \simeq 0.633$,
 there are two independent exponents in this model, compared to only one in the directed
 polymer model.

Huse and Henley [1] refer to the possibility that in the directed polymer
 model the random numbers are taken from a bimodal  probability distribution,
 and stated that the roughness exponent has the same value as in the continuous case, i.e. 
$\nu = 2/3$.  On
 the other hand, more recently, Lebedev and Zhang [6] claimed that in the bimodal
 case, for all $p \leq p_{c}$   the exponent is actually the same as in the
 directed percolation case, i.e.  $\nu \simeq 0.633$.

 In this Letter we address  this controversial issue by studying the directed polymer
  model with  bimodal (0,1) distribution. The numerical simulations lead to the following
  main conclusions:
\begin{enumerate}
\item For $p < p_{c}$ and for $p > p_{c}$, for large t the value of the roughness exponent is
 $ \simeq 2/3$, 
the directed polymer value.
 
\item For $p = p_{c}$ , the value of the roughness exponent is $ \simeq 0.633$, 
 the directed percolation value.

\item For $p < p_{c}$  and $t \gg L$, the sum of the random numbers along the optimal
 path is inversely proportional to the mean directed percolation cluster length,
i.e.  $E\sim t/L$. 
\end{enumerate}

  It should be emphasized that a study of the directed polymer model yields
   dependence on both the critical probability and mean percolation cluster 
    length, L,  of directed percolation. So the general conclusion is that
     the two models are closely \vspace{0.4in} related.

 In this study the random numbers are taken from a bimodal (0,1) distribution, 
 and thus there is usually more than one optimal path and more than one optimal endpoint.
 If there exists a real percolation cluster spanning from the origin to the base, 
 all the optimal endpoints (and no other point of the base) belong to this percolation
 cluster, but otherwise, the optimal paths and endpoints are not necessarily connected 
 in any special manner. The set of optimal endpoints can be characterized by  
  two  variables:
  
  \begin{enumerate}
  \item $D_{0}$ - the distance of the center of the optimal endpoints set from the
  center of the base. The center of the set is computed by $(X_{l} + X_{r})/2$ 
  where $X_{l}$ and $X_{r}$ are the leftmost and rightmost points of the set.
  \item $W_{0} $ - the width of the optimal endpoints set,
   which is computed by
  \(W_{0} = (X_{r} - X_{l})\).
 \end{enumerate}

 It is expected that for large t, $D_{0} \sim t^{\nu_{D}}$, and $W_{0} \sim t^{\nu_{W}}$,
 where $\nu_{D}$ and $\nu_{W}$ are in the range [0,1]. In order to test this
 hypothesis and estimate $\nu_{D} $ and $\nu_{W}$, we define the local roughness exponent
 of a variable V at the point t, as the slope of the curve relating the logarithm of
 V to the logarithm of t. The numerical simulations provide estimates for the local
 exponents of $D_{0}$ and $W_{0}$ by computing
 $(\log V(2t) - \log V(t/2))/\log 4$, where V is either $D_{0}$ or $W_{0}$. These estimates
 are presented in Figures 1 and 2 for 5 probabilities: $p = 0.25$, $p = 0.5$,
 $p = 0.62$, $p = 0.64$ and $p = 0.6447 (\simeq p_{c})$. 
  In Figure 1  it is seen  that the slope
 of the $p \simeq p_{c} $ curve is stable at a value $\simeq 0.63$, while as p is 
 further below $p_{c} $,  the slope seems to crossover from $\simeq 0.63$ to a value
 close to 2/3. 
   In Figure 2 it is  seen  that the slope of the $p \simeq p_{c} $ curve converges to a value
 $\simeq$ 0.63, while for $p < p_{c}$ the slope declines 
   and approaches  (for the probabilities further
  from $p_{c}$)  a value close to 1/3. These crossovers from directed percolation exponents
 to directed polymer exponents can be explained due to the relation between the triangle 
 size t and the longitudinal correlation length $\xi_{\parallel}$ [2]. For $t < \xi_\parallel$
 we  expect properties of directed percolation, while for $t > \xi_{\parallel}$ we expect 
 properties of directed polymer.

  The results presented in Figure 2 deserve some explanation, as one might expect
   that $W_{0}$  should be proportional to W - the width of a typical directed 
  percolation cluster (in the relevant probability). If this was the case, we would
  get in Figure 2 an initial slope of 0.63, and than the slope would decline to
  zero. Evidently, the picture is different, as  for each probability 
  $p < p_{c}$ the slope will eventually converge to a value close to 1/3.
 The reason for this lies in the ultrametric tree structure of the 
   directed polymer problem. It was shown [7]
    that in the continuous distribution  case, the region of endpoints of paths
   whose value  difference from the value of the  optimal path is smaller than a (small) constant, 
   increases as $t^{1/3}$. In our case of discrete  values, the same should hold 
   for zero  difference between the values of the optimal paths.

 The main conclusion from Figures 1,2 is that for $p < p_{c}$ and sufficiently  large
 t, the optimal endpoints set of width $\sim t^{1/3}$ is located around a point 
 whose distance from the center is $\sim t^{2/3}$. As for $t \gg 1$, $ t^{1/3}$ is negligible
 compared to $t^{2/3}$, the situation is similar to the continuous distribution case,
 where instead of one endpoint there is a "cloud" of endpoints whose distance from
 the center is $\sim t^{2/3}$.

 For $p > p_{c}$
 there is a finite probability for points of the base to belong to a real directed
 percolation cluster, so that $W_{0} \sim t$. On the other hand, we find numerically that 
 $D_{0} \sim t^{1/2}$. As $W_{0} \gg D_{0}$, it is certain that members of the optimal
 endpoints set are found in both sides of  the center, and another  definition of the
 distance is needed. A definition  that takes into account the fact that the directed polymer
 is equally likely to choose any one of the optimal paths, is 
 $D_{r} = \sum_{i} n_{i}|x_{i}|/\sum_{i}n_{i}$,  where $n_{i}$ is the number of optimal 
  paths  whose endpoint is $x_{i}$. The local exponents of $D_{r}$ 
 are presented in Figure 3 for 3 probabilities: $p = 0.5$, $p = 0.6447 (\simeq p_{c})$,
 and $p = 0.75$.

 As might be expected, the results for $p < p_{c}$ and $p \simeq p_{c}$ approach 
 the values 2/3 and 0.63 respectively, further supporting  that for $p < p_{c}$, $ \nu = 2/3$,
 while for $p = p_{c}$, $  \nu \simeq 0.63$. The results for $p>p_{c}$ are inconclusive,
 as the local exponent does not yet converge at this triangle size. But, for
  $p > p_{c}$  it is expected that the situation is governed by one big percolation
 cluster. If this is the case, it is
 possible (and much less compuer power demanding) to "grow" percolation clusters and to
 find the dependence of $D_{r}$ on t in that way. The local exponents of $D_{r}$ for
 directed polymer at $p = 0.75$ and for directed percolation at the same probability
 are shown in figure 4.

  As can be seen in figure 4, the results of directed polymer and directed percolation
 are statistically indistinguishable, and it is quite safe to assume that this situation
 will not change for larger t. The local exponent for directed percolation 
 approches a value $\simeq 2/3$, a result which was obtained earlier by Balents and 
 Kardar [8]. Thus we conclude that for $p > p_{c}$, $\nu$ of directed polymer
has the same value as for
 $p < p_{c}$, i.e.  $\nu \simeq 2/3$.

 It was mentioned above that the results of Figures 1,2 can be explained in terms of the
 longitudinal correlation length $\xi_{\parallel}$. More direct result follows from a picture
 of the optimal path as a series of zero sum segments
 whose mean length is L, connected by single bonds of value 1. (Obviously, this picture
 holds only for $t \gg  L$). According to  this picture,we expect that  $E \sim t/L$, and for
 fixed t, $E \times L \sim $ constant. Figure 5 presents the results for $E/E_{0}$, $L/L_{0}$,
 and $(E \times L)/(E_{0} \times L_{0})$   in the range of probabilities $0.5 < p < 0.64$,  where $E_{0}$
 is E at $p = 0.5$ and $L_{0}$ is L at $p = 0.5$. 

As can be seen in Figure 5, E and L form mirror reflection of each other over two orders of magnitude,
 while $E \times L$ is a slightly decreasing function of p.  So there is almost one to one correspondence
 between results obtained from the directed polymer model (the values of E), and results obtained  
 \emph{independently} from the directed percolation \vspace{0.3in} model (the values of L).

    In conclusion, it was shown that the bimodal distribution directed polymer model
  can be characterized in terms of the directed percolation model, and that
     the percolation thresehold probability $p_{c}$ plays a critical role in 
  the directed polymer case. 
   
\newpage

 \begin{center}
{\bfseries REFERENCES}
\end{center}

     \begin{enumerate}

\item D.A. Huse and C.L. Henley, Phys. Rev. Lett. 54, 2708 (1985).  

\item W. Kinzel in: Percolation Structures and Processes, ed. G. Deutscher,
 R. Zallen and J. Adler, 1983 (Hilger, Bristol).

\item J. Kert\'{e}sz, V.K. Horv\'{a}th and F. Weber, Fractals, 1, 67 (1992).

\item Fractals and Disordered Systems, ed. A. Bunde and S. Havlin, 2nd Edition, 
 1995 (Springer, Berlin).

\item A.-L. Barab\'{a}si and H.E. Stanley, Fractal Concepts in Surface Growth, 1995 
 (Cambridge University Press, Cambridge).

\item N.I. Lebedev and Y.-C. Zhang, J. Phys. A, 28, L1-L6, (1995).

\item E. Perlsman and M. Schwartz, Europhys. Lett. 17, 11, (1992).

\item L. Balents and M. Kardar, J. Stat. Phys. 67, 1, (1992).
      \end{enumerate}                            

%\vspace{0.9in}

\vfill\break

%\begin{center}
%{\bfseries FIGURES}
%\end{center}
%\vspace{0.2in}

\begin{figure}
\epsfig{file=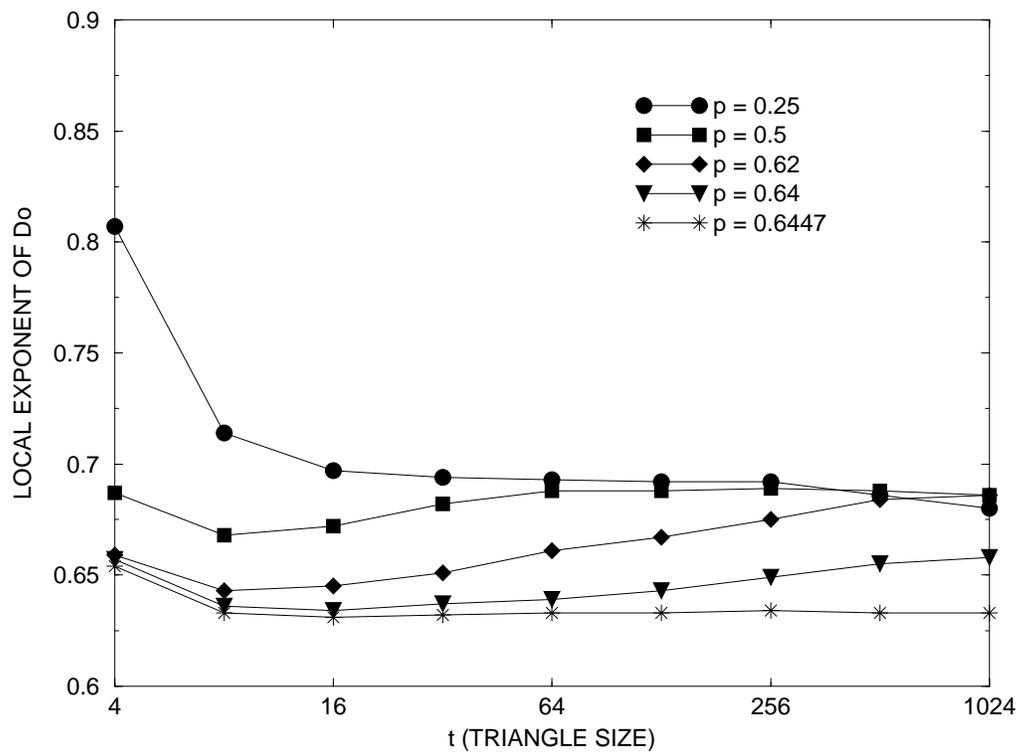,width=10cm,clip=,bbllx=95,bblly=65,bburx=550,bbury=680,angle=-90}
\caption{%Fig. 1.
Plot of the local exponent of $D_{0}$ as a function of t for several values of p.
}
\end{figure}
\vspace{0.2in}

\begin{figure}
\epsfig{file=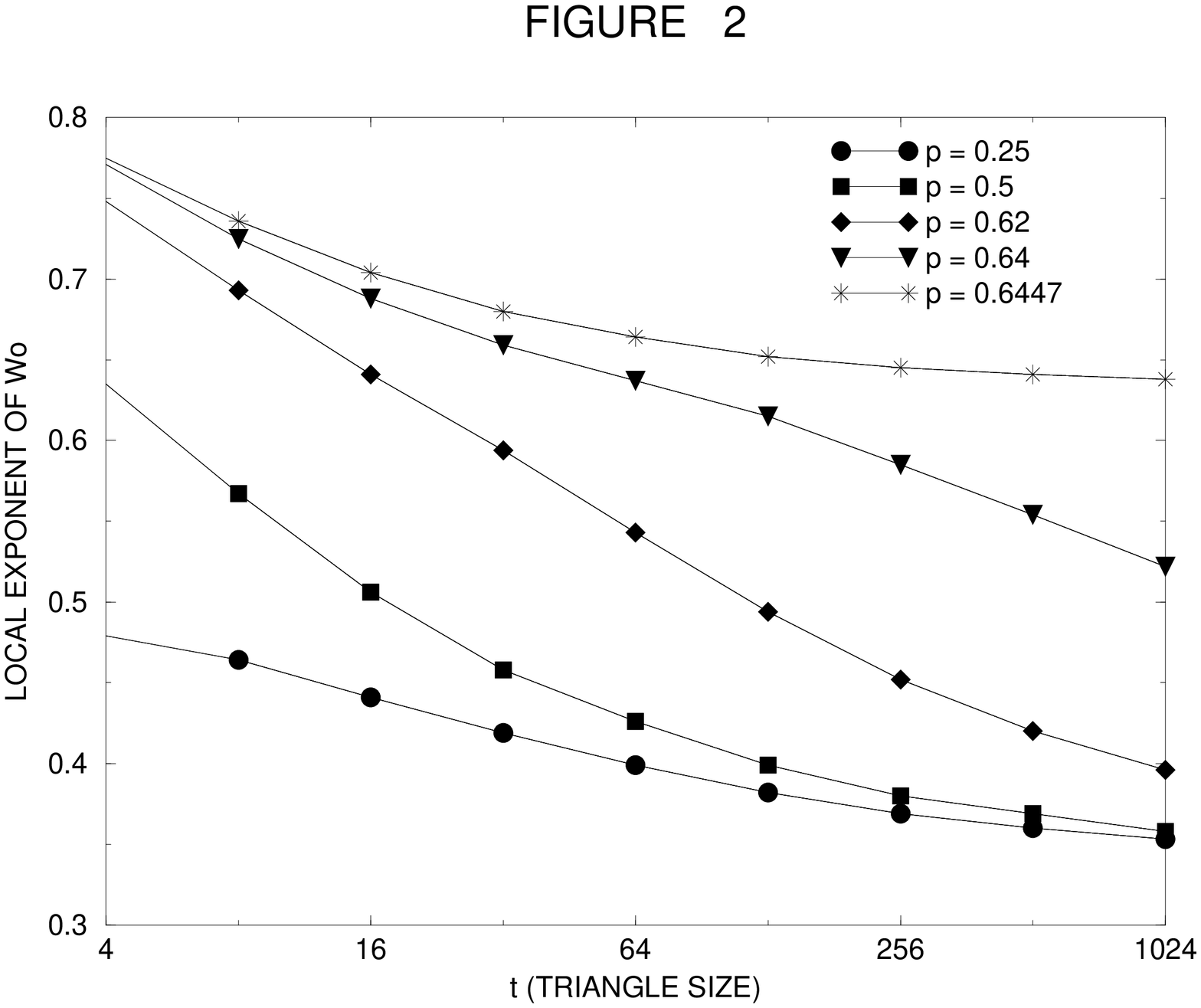,width=10cm,clip=,bbllx=95,bblly=75,bburx=550,bbury=680,angle=-90}
\caption{%Fig. 2.
Plot of the local exponent of $W_{0}$ as a function of t for several values of p. 
}
\end{figure}
\vspace{0.2in}

\begin{figure}
\epsfig{file=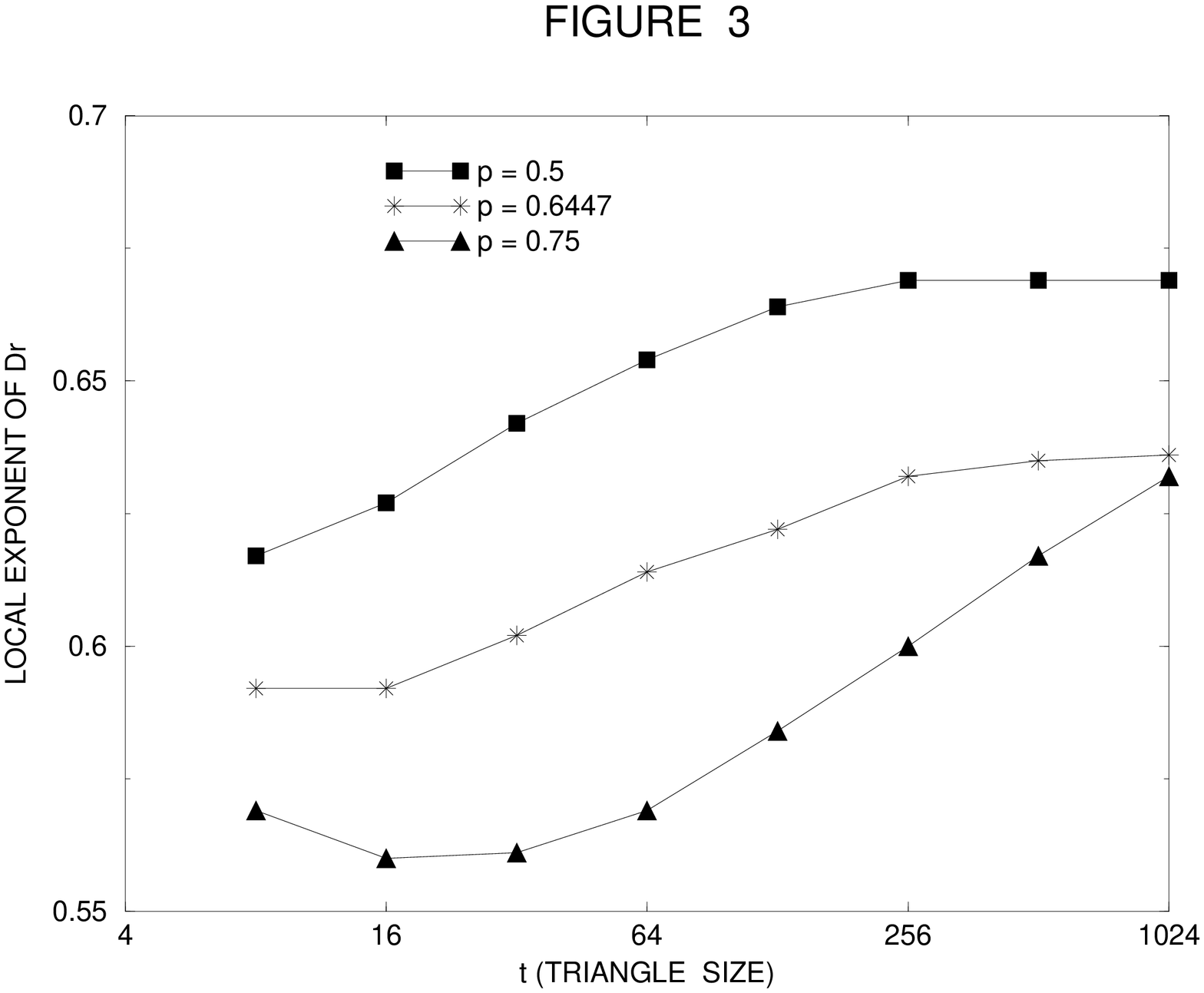,width=10cm,clip=,bbllx=95,bblly=65,bburx=550,bbury=680,angle=-90}
\caption{%Fig. 3.
Plot of the local exponent of $D_{r}$ as a function of t for several values of p.
}
\end{figure}
\vspace{0.2in}

\begin{figure}
\epsfig{file=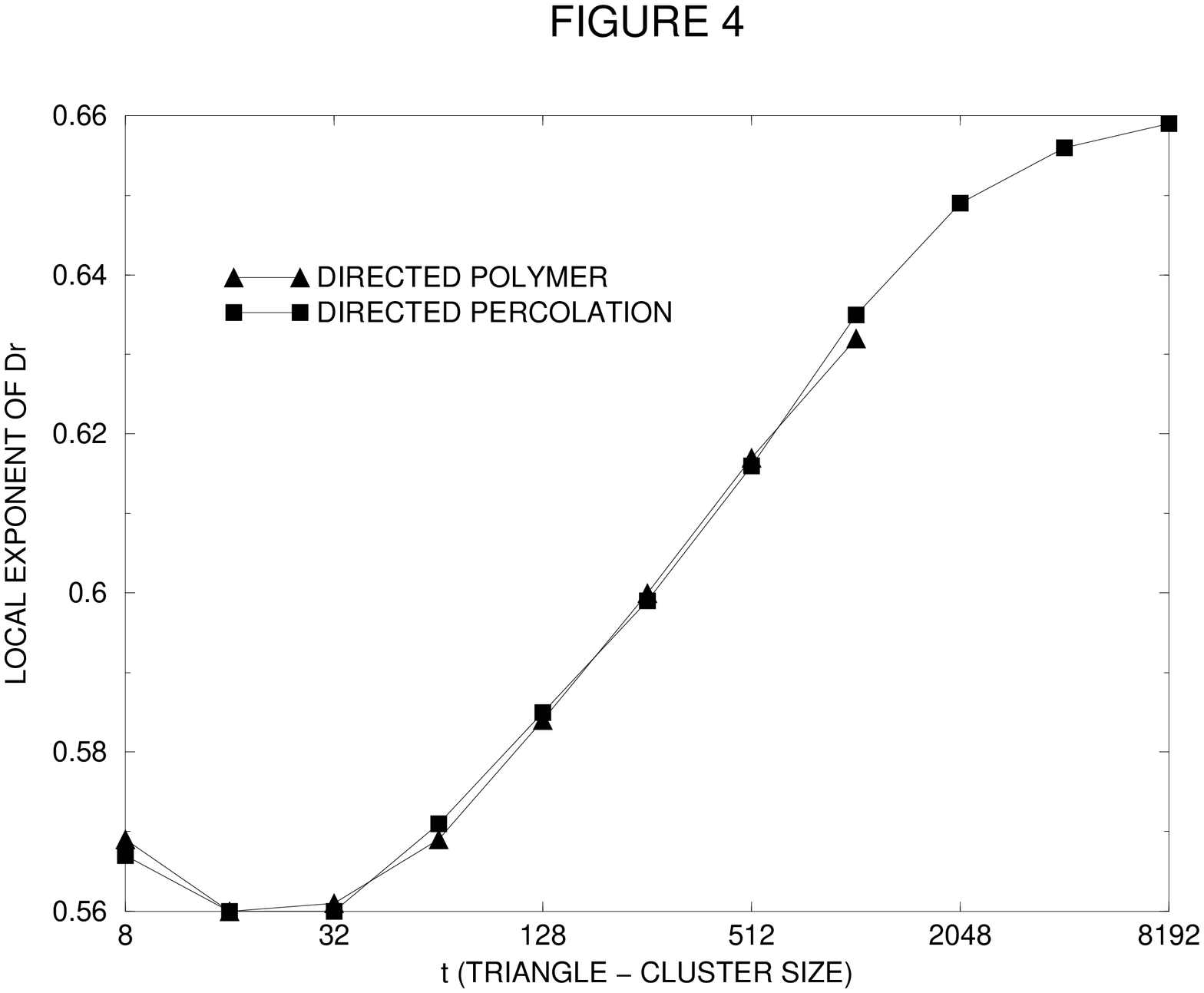,width=10cm,clip=,bbllx=95,bblly=65,bburx=550,bbury=680,angle=-90}
\caption{%Fig. 4.
Plot of the local exponent of $D_{r}$ as a function of t for directed polymer
and directed percolation at $p = 0.75$.\vspace{0.2in}
}
\end{figure}
\vspace{0.2in}

\begin{figure}
\epsfig{file=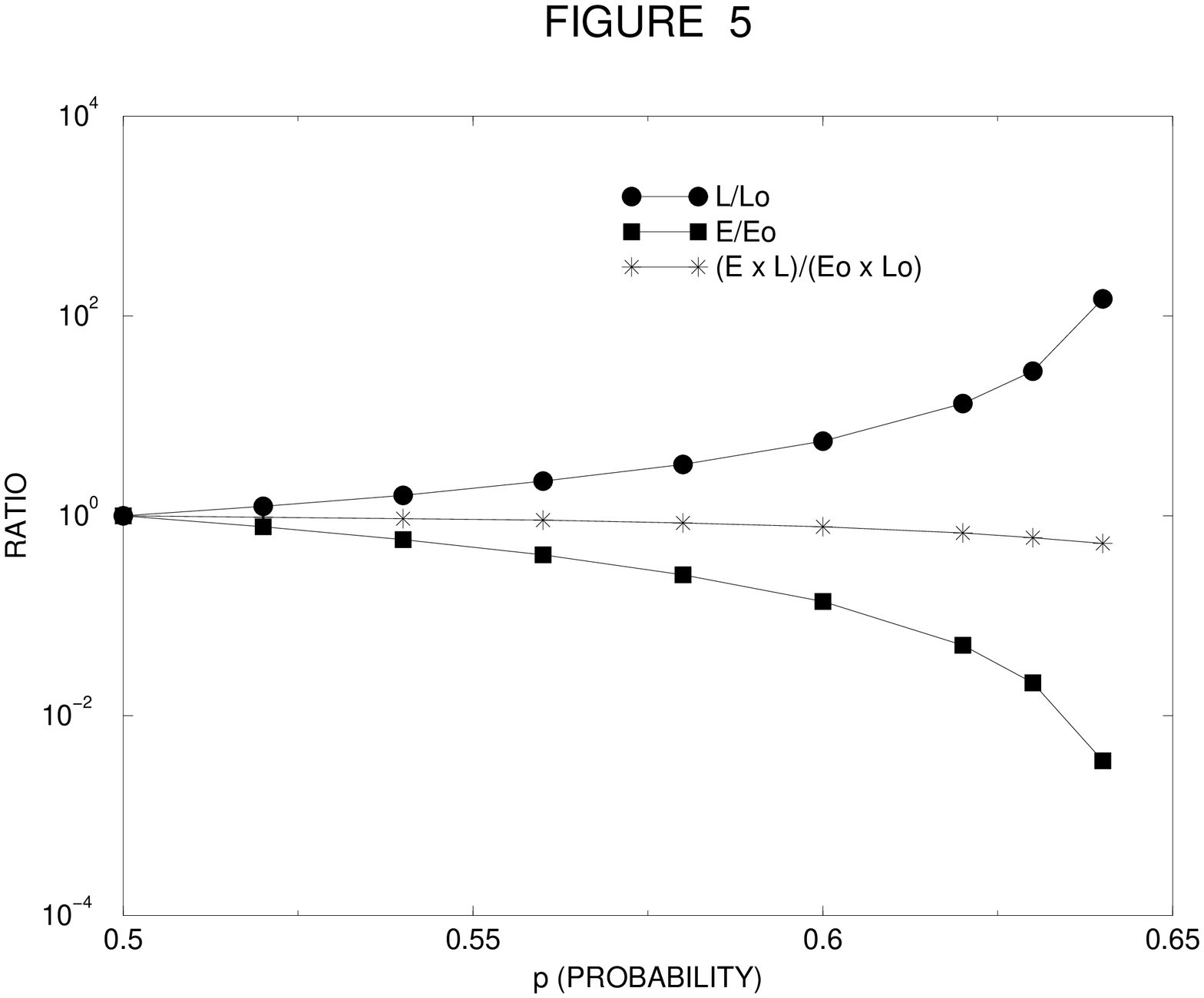,width=10cm,clip=,bbllx=95,bblly=65,bburx=550,bbury=680,angle=-90}
\caption{%Fig. 5.
Plot of $E/E_{0}$, $L/L_{0}$, and $(E \times L)/(E_{0} \times L_{0})$ as a function
of the probability p. $E_{0}$ and $L_{0}$ are the values at $p = 0.5$.
}
\end{figure}

\end{document}